\documentstyle[preprint,aps]{revtex}
\begin{document}
\author{Jian-Qi Shen $^{1,}$$^{2}$ \footnote{E-mail address: jqshen@coer.zju.edu.cn}}
\address{$^{1}$  Centre for Optical
and Electromagnetic Research, State Key Laboratory of Modern
Optical Instrumentation, Zhejiang University,
Hangzhou Yuquan 310027, P.R. China\\
$^{2}$Zhejiang Institute of Modern Physics and Department of
Physics, Zhejiang University, Hangzhou 310027, P.R. China}
\date{\today }
\title{Optical Refractive Index of Massive Particles and Physical Meanings of Left-handed Media}
\maketitle

\begin{abstract}
In this Letter the expression for the refractive index of de
Broglie wave in the presence of a potential field is obtained and
based on this, the physical meanings of negative index of
refraction is revealed. We demonstrate that the electromagnetic
wave propagation in a left-handed medium with negative refractive
index behaves just like that of {\it antiphotons}, which is
required of the complex vector field theory. It is believed that
the complex vector field theory is helpful in considering the wave
propagation and photonic band gap structure in the left-handed
medium photonic crystals with a periodicity in negative and
positive index of refraction.
\end{abstract}

More recently, a kind of artificial composite metamaterials (the
so-called {\it left-handed media}) having a frequency band where
the effective permittivity and the effective permeability are
simultaneously negative attracts considerable attention of many
authors both experimentally and
theoretically\cite{Smith,Klimov,Shelby,Ziolkowski2,Kong,Garcia,Jianqi,Shen}.
In 1967\footnote{Note that, in the literature, some authors
mentioned the wrong year when Veselago suggested the {\it
left-handed media}. They claimed that Veselago proposed or
introduced the concept of {\it left-handed media} in 1968 or 1964.
On the contrary, the true history is as follows: Veselago's
excellent paper was first published in Russian in July, 1967 [Usp.
Fiz. Nauk {\bf 92}, 517-526 (1967)]. This original paper was
translated into English by W.H. Furry and published again in 1968
in the journal of Sov. Phys. Usp.\cite{Veselago}. Unfortunately,
Furry stated erroneously in his English translation that the
original version of Veselago' work was first published in 1964.},
Veselago first considered this peculiar medium and showed from
Maxwellian equations that such media having negative
simultaneously negative $\epsilon $ and $\mu $ exhibit a negative
index of refraction, {\it i.e.}, $n=-\sqrt{\epsilon \mu
}$\cite{Veselago}. It follows from the Maxwell's curl equations
that the phase velocity of light wave propagating inside this
medium is pointed opposite to the direction of energy flow, that
is, the Poynting vector and wave vector of electromagnetic wave
would be antiparallel, {\it i.e.}, the vector {\bf {k}}, the
electric field {\bf {E}} and the magnetic field {\bf {H}} form a
left-handed system; thus Veselago referred to such materials as
``left-handed'' media, and correspondingly, the ordinary medium in
which {\bf {k}}, {\bf {E}} and {\bf {H}} form a right-handed
system may be termed the ``right-handed'' one. Other authors call
this class of materials ``negative-index media
(NIM)''\cite{Gerardin}, ``backward media (BWM)''\cite{Lindell},
``double negative media (DNM)''\cite{Ziolkowski2} and Veselago's
media. There exist a number of peculiar electromagnetic and
optical properties, for instance, many dramatically different
propagation characteristics stem from the sign change of the
optical refractive index and phase velocity, including reversal of
both the Doppler shift and Cherenkov radiation, anomalous
refraction, amplification of evanescent waves\cite{Pendryprl},
unusual photon tunneling\cite{Zhang}, modified spontaneous
emission rates and even reversals of radiation pressure to
radiation tension\cite{Klimov}. In experiments, this artificial
negative electric permittivity media may be obtained by using the
{\it array of long metallic wires} (ALMWs)\cite{Pendry2}, which
simulates the plasma behavior at microwave frequencies, and the
artificial negative magnetic permeability media may be built up by
using small resonant metallic particles, {\it e.g.}, the {\it
split ring resonators} (SRRs), with very high magnetic
polarizability\cite{Pendry1,Pendry3,Maslovski}. A combination of
the two structures yields a left-handed medium. Recently, Shelby
{\it et al.} reported their first experimental realization of this
artificial composite medium, the permittivity and permeability of
which have negative real parts\cite{Shelby}. One of the potential
applications of negative refractive index materials is to
fabricate the so-called ``superlenses'' (perfect lenses):
specifically, a slab of such materials may has the power to focus
all Fourier components of a 2D image, even those that do not
propagate in a radiative manner\cite{Pendryprl,Hooft}.

A number of novel electromagnetic and optical properties in
left-handed media result from the negative index of refraction.
But what is its physical meanings? Although in the literature many
researchers have investigated the negative refractive index of
left-handed media by a variety of means of applied
electromagnetism, classical optics, materials science as well as
condensed matter
physics\cite{Smith,Klimov,Shelby,Ziolkowski2,Kong,Garcia,Jianqi},
less attention than it deserves is paid to the physical meanings
of negative refractive index. In this Letter we will study this
fundamental problem from the purely physical point of view.

Before we consider the negative refractive index of left-handed
media, we first deal with the ``optical refractive index" problem
of de Broglie wave, the results of which will be helpful in
discussing the former problem.

For the case of de Broglie particle moving in a force field, we
can also consider its ``optical refractive index" $n$. According
to the Einstein- de Broglie relation, the dispersion relation of
the de Broglie particle with rest mass $m_{0}$ in a scalar
potential field $V\left( {\bf x}\right) $ agrees with
\begin{equation}
\left( \omega -\phi \right)
^{2}=k^{2}c^{2}+\frac{m_{0}^{2}c^{4}}{\hbar ^{2}}, \label{eq1}
\end{equation}
where $\phi$ is defined to be $\phi=\frac{V}{\hbar }$, and $k$,
$\hbar$ and $c$ stand for the wave vector, Plank constant and
speed of light in a vacuum, respectively. It follows from
Eq.(\ref{eq1}) that
\begin{equation}
\omega ^{2}\left[ 1-\frac{m_{0}^{2}c^{4}}{\hbar ^{2}\omega
^{2}}-2\frac{\phi }{\omega }+\left( \frac{\phi }{\omega }\right)
^{2}\right] =k^{2}c^{2},     \label{eq2}
\end{equation}
which yields
\begin{equation}
\frac{\omega
^{2}}{k^{2}}=\frac{c^{2}}{1-\frac{m_{0}^{2}c^{4}}{\hbar ^{2}\omega
^{2}}-2\frac{\phi }{\omega }+\left( \frac{\phi }{\omega }\right)
^{2}}.   \label{eq3}
\end{equation}
Compared Eq.(\ref{eq3}) with the dispersion relation $\frac{\omega
^{2}}{k^{2}}=\frac{c^{2}}{n^{2}}$, one can arrive at
\begin{equation}
n^{2}=1-\frac{m_{0}^{2}c^{4}}{\hbar ^{2}\omega ^{2}}-2\frac{\phi
}{\omega }+\left( \frac{\phi }{\omega }\right) ^{2}, \label{eq4}
\end{equation}
which is the square of ``optical refractive index" of de Broglie
wave in the presence of a potential field $\phi $.

It is of physical interest to discuss the Fermat's principle of de
Broglie wave in a potential field $V$. One can readily verify that
the variation $\delta \int cn^{2}{\rm d}t=0$ can serve as the
mathematical expression for the Fermat's principle of de Broglie
particle (for the derivation of this expression, the readers may
be referred to the Appendix). If, for example, by using the
weak-field and low-motion approximation where $\left| \frac{\phi
}{\omega }\right| \ll 1$, $\frac{v^{2}}{c^{2}}\ll 1$ and $\left(
\frac{\phi }{\omega }\right) ^{2}$ in Eq.(\ref{eq4}) can therefore
be ignored, $1-\frac{m_{0}^{2}c^{4}}{\hbar ^{2}\omega ^{2}}\simeq
\frac{v^{2}}{c^{2}}$, then $n^{2}$ is approximately equal to
\begin{equation}
n^{2}\simeq \frac{2}{mc^{2}}\left( \frac{1}{2}mv^{2}-V\right).
\label{eq5}
\end{equation}
Thus in the non-relativistic case, Fermat's principle, $\delta
\int cn^{2}{\rm d}t=0$, is equivalent to the following well-known
action principle $\delta \int L({\bf x},{\bf v}){\rm d}t=0$
(differing only by a constant coefficient $\frac{2}{mc^{2}}$),
where the function $L({\bf x},{\bf v})=\frac{1}{2}mv^{2}-V({\bf
x})$ denotes the Lagrangian of massive particle in a potential
field $V({\bf x})$. Hence the above formulation for the ``optical
refractive index" of the de Broglie particle inside a potential
field is said to be self-consistent.

According to Eq.(\ref{eq4}), there are two square roots ({\it
i.e.}, positive and negative roots, $n_{+}$, $n_{-}$) for the
refractive index of de Broglie wave. Indeed, it is verified that
the particle and its antiparticle possess these two square roots,
respectively. If the refractive index of particle is positive,
then that of its antiparticle will acquire a minus sign, which may
be seen by utilizing the charge conjugation transformation:
$\omega\rightarrow-\omega$, $\phi\rightarrow-\phi$. If, for
example, we consider the refractive index of electron and positron
by regarding the positron as the negative energy solution to
Dirac's equation, then we can classify the solutions of Dirac's
equation into the following four cases:

(i) \quad  $E=E_{+}$, \quad $h=\hbar k$, \quad $n=n_{+}$;

(ii) \quad  $E=E_{+}$, \quad $h=-\hbar k$, \quad $n=n_{+}$;

(iii) \quad $E=E_{-}$, \quad $h=\hbar k$,\quad  $n=n_{-}$;

(iv) \quad   $E=E_{-}$,  \quad  $h=-\hbar k$, \quad   $n=n_{-}$,
\\
where $h$, $E_{+}$ and
$E_{-}$ respectively represent the helicity, positive and negative
energy corresponding to the electron and positron. It is apparent
that the relationship between the cases (i) and (iv) is just the
charge conjugation transformation. This transformation also
relates the case (ii) to (iii).

Compare the case of wave propagation in left-handed media with the
results presented above, we can conclude that the negative optical
refractive index in left-handed media corresponds to the
antiparticles of photons (the so-called {\it antiphotons}).
However, as is well known, there exist no {\it antiphotons} in
free space. The theoretical reason for this is that the
four-dimensional vector potentials $A_{\mu}$ with $\mu=0,1,2,3$
are always taking the real numbers. But in a dispersive and
absorptive medium, as an effective medium theory, the vector
potentials $A_{\mu}$ of which may probably take the complex
numbers. Such complex vector field theory has been considered
previously\cite{Lurie}. Here the Lagrangian density is written
${\mathcal L}=-\frac{1}{2}F^{\ast}_{\mu\nu}F_{\mu\nu}$, where the
electromagnetic field tensors
$F_{\mu\nu}=\partial_{\mu}A_{\nu}-\partial_{\nu}A_{\mu}$,
$F^{\ast}_{\mu\nu}=\partial_{\mu}A^{\ast}_{\nu}-\partial_{\nu}A^{\ast}_{\mu}$
with $A^{\ast}_{\mu}$ being the complex conjugation of $A_{\mu}$.
The complex four-dimensional vector potentials $A_{\mu}$ and
$A^{\ast}_{\mu}$ describe the propagating behavior of both photons
and antiphotons.

Now a problem left to us is that does the complex vector field
theory apply well in investigating the light propagation inside
the negative refractive index medium. Detailed analysis will show
that this theory is truly applicable to the consideration of
optical index of refraction and wave propagation in left-handed
media. This will be proved in what follows by using another
mathematical form
\begin{equation}
\frac{i}{c_{n}}\frac{\partial}{\partial t}{\bf M}=\nabla\times{\bf
M},  \quad       \nabla\cdot{\bf M}=0         \label{eq6}
\end{equation}
for the Maxwellian equations, where ${\bf M}=\sqrt{\epsilon}{\bf
E}+i\sqrt{\mu}{\bf H}$, $c_{n}=1/\sqrt{\epsilon\mu}$, $\epsilon$
and $\mu$ are absolute dielectric constant and magnetic
conductivity of the medium, respectively. It can be easily
verified that the Maxwellian equations can be rewritten as
Eq.(\ref{eq6}).

If $\epsilon>0$ and $\mu>0$ are assumed, then the light
propagation inside the isotropic linear right-handed medium
(regular medium) is governed by Eq.(\ref{eq6}). For the
time-harmonic electromagnetic wave in the right-handed medium,
Eq.(\ref{eq6}) can be rewritten
\begin{equation}
i\frac{\omega}{c_{n}}{\bf M}=-{\bf k}\times{\bf M} \label{eq7}
\end{equation}
with $\omega$ being the frequency of electromagnetic wave.

Now let us take the complex conjugation of the two sides of
Eq.(\ref{eq7}), and the result is of the form
\begin{equation}
i\frac{\omega}{c_{n}}{\bf M^{\ast}}={\bf k}\times{\bf M^{\ast}},
\label{eq8}
\end{equation}
where the field ${\bf M^{\ast}}$ is expressed in terms of both
electric and magnetic fields, ${\bf E^{\ast}}$ and ${\bf
H^{\ast}}$, {\it i.e.}, ${\bf M^{\ast}}=\sqrt{\epsilon}{\bf
E^{\ast}}-i\sqrt{\mu}{\bf H^{\ast}}$. Thus, we have
\begin{equation}
i\frac{\omega}{c_{n}}[\sqrt{\epsilon}{\bf
E^{\ast}}-i\sqrt{\mu}{\bf H^{\ast}}]={\bf
k}\times[\sqrt{\epsilon}{\bf E^{\ast}}-i\sqrt{\mu}{\bf H^{\ast}}].
\label{eq9}
\end{equation}
Multiplying the two sides of Eq.(\ref{eq9}) by the imaginary unit
$i$, one can arrive at
\begin{equation}
i\frac{\omega}{c_{n}}[\sqrt{-\epsilon}{\bf
E^{\ast}}-i\sqrt{-\mu}{\bf H^{\ast}}]={\bf
k}\times[\sqrt{-\epsilon}{\bf E^{\ast}}-i\sqrt{-\mu}{\bf
H^{\ast}}], \label{eq10}
\end{equation}
where the magnitude of $c_{n}$ ({\it i.e.},
$1/\sqrt{\epsilon\mu}$) does not change but it can be rewritten as
a new form $c_{n}=1/\sqrt{(-\epsilon)(-\mu)}$. According to
Eq.(\ref{eq10}), one can obtain the following relations
\begin{equation}
{\bf k}\times{\bf E^{\ast}}=\omega(-\mu){\bf H^{\ast}},   \quad
{\bf k}\times{\bf H^{\ast}}=-\omega(-\epsilon){\bf E^{\ast}}.
\label{eq11}
\end{equation}
Note that here $\epsilon>0$, $\mu>0$. Thus it follows from
Eq.(\ref{eq11}) that the wave vector ${\bf {k}}$, the electric
field ${\bf {E^{\ast}}}$ and the magnetic field ${\bf {H^{\ast}}}$
of the electromagnetic wave in this medium, through which the
fields ${\bf E^{\ast}}$ and ${\bf H^{\ast}}$ propagate, form a
left-handed system. This, therefore, means that here the Poynting
vector ${\bf S}$ is pointed opposite to the direction of the wave
vector of electromagnetic wave. For this reason, it is expected
that the reversal of wave vector, instantaneous helicity
inversion\cite{Shenpla} and anomalous refraction will take place
when an incident lightwave from the right-handed medium travels to
the interfaces between left- and right- handed media. Since the
direction of phase velocity and energy flow of lightwave
propagating in the left-handed medium would be antiparallel, the
change of wave vector ${\bf k}$ will truly occur during the light
propagation through the interfaces. The inversion of ${\bf k}$
means that its magnitude acquires a minus sign, namely, the
optical index of refraction $n$ becomes negative since the
magnitude of ${\bf k}$ is defined to be $k=n\omega/c$. In view of
the above discussion, it is concluded that Eq.(\ref{eq10}) governs
the propagation of time-harmonic planar electromagnetic wave,
where the electromagnetic fields are described by ${\bf E^{\ast}}$
and ${\bf H^{\ast}}$.

Thus by making use of the complex vector field theory we obtain a
formulation which can treat the wave propagation in both right-
and left- handed media. One of the advantages of the present
formulation is that the minus sign of negative refractive index
can be placed into the theoretical calculations by using the above
unified approach rather than {\it by hand}. Such unified approach
will be useful in dealing with the electromagnetic wave
propagation and photonic band gap structures in the so-called
left-handed medium photonic crystals, which are artificial
materials patterned with a periodicity in negative and positive
index of refraction.

To summarize, we study the ``optical refractive index" of de
Broglie wave and reveal the physical meanings of negative
refractive index in left-handed media. It is shown that the
electromagnetic wave propagation in negative refractive index
media behaves like that of {\it antiphotons}, which requires that
we should take into account the complex vector field theory. This
will provide us with a unified formulation to treat the wave
propagation in both left- and right- handed media.
\\ \\
\textbf{Acknowledgements}  This project was supported by the
National Natural Science Foundation of China under the project No.
$90101024$. The author is grateful to S.L. He for useful
discussion on the wave propagation in left-handed media.
 \\ \\
{\bf Appendix}

In this Appendix, we will discuss the Fermat's principle of de
Broglie wave. First we consider the relativistic free-particle
case where the Einstein-de Broglie relation $(\hbar
k)^{2}c^{2}=(\hbar \omega)^{2}-{m_{0}^{2}c^{4}}$ is utilized. By
applying the dispersion relation $\frac{\omega
^{2}}{k^{2}}=\frac{c^{2}}{n^{2}}$ to the Einstein-de Broglie
relation, one can arrive at
\begin{equation}
\frac{m_{0}^{2}c^{4}}{\hbar ^{2}}=\left(1-n^{2}\right) \omega ^{2}
\eqnum{A1}. \label{eqA1}
\end{equation}
and consequently derive
\begin{equation}
\hbar \omega =\frac{m_{0}c^{2}}{\sqrt{1-n^{2}}}. \eqnum{A2}
\label{eqA2}
\end{equation}

Compared Eq.(\ref{eqA2}) with the relativistic energy-velocity
formula (with $v$ being the particle velocity)
\begin{equation}
\hbar \omega =\frac{m_{0}c^{2}}{\sqrt{1-(\frac{v}{c})^{2}}},
\eqnum{A3} \label{eqA3}
\end{equation}
we obtain $n=\frac{v}{c}$. According to the original expression
for the Fermat's principle, {\it i.e.}, $\delta \int n{\rm d}l=0$,
where $n{\rm d}l$ is the ``optical" path element with ${\rm
d}l=v{\rm d}t=nc{\rm d}t$, the Fermat's principle for the de
Broglie particle can be rewritten
\begin{equation}
\delta \int \frac{v}{c}v{\rm d}t=0 \quad          {\rm or} \quad
\delta\int cn^{2}{\rm d}t=0. \eqnum{A4} \label{A4}
\end{equation}
 In order to consider the wave propagation of the Broglie particle
 in a potential field, we assume that the latter formula ({\it i.e.}, $\delta\int cn^{2}{\rm d}t=0$) in
 Eq.(\ref{A4}) is still valid and would also apply very well in this case. It follows from
 Eq.(\ref{eq5})that this assumption truly holds at least for the
 non-relativistic case.

\end{document}